\documentclass[twocolumn,prb,notitlepage,floats,superscriptaddress,amsmath,amssymb]{revtex4-2}

\usepackage[version=3]{mhchem} 
\usepackage{bm}
\usepackage[utf8]{inputenc}
\usepackage[T1]{fontenc}
\usepackage{graphicx}
\usepackage{upgreek}
\usepackage{color}
\usepackage{ulem}
\usepackage{hyperref} 

\hypersetup{colorlinks, citecolor=blue, filecolor=blue ,linkcolor=blue , urlcolor=blue, pdftex}
\usepackage{sidecap}
\usepackage{amssymb}
\usepackage{amsmath}
\usepackage[caption=false]{subfig}
\usepackage{multirow}

\def \FUW{University of Warsaw, Faculty of Physics, 02-093 Warsaw, Poland}
\def \Prague{Charles University, Faculty of Mathematics and Physics,  CZ-121 16 Prague, Czech Republic}
\def \Watanabe{Research Center for Electronic and Optical Materials, National Institute for Materials Science, 1-1 Namiki, Tsukuba 305-0044, Japan}
\def \Taniguchi{Research Center for Materials Nanoarchitectonics, National Institute for Materials Science,  1-1 Namiki, Tsukuba 305-0044, Japan}

\begin{document}

\title{Doping-Induced Brightening of Dark Excitons and Trions in a WSe$_2$ Monolayer}

\author{Grzegorz Krasucki}
\email{grzegorz.krasucki@fuw.edu.pl}
\affiliation{\FUW}
\author{Artur O. Slobodeniuk}
\affiliation{\Prague}
\author{Kacper Walczyk}
\affiliation{\FUW}
\author{Katarzyna Olkowska-Pucko}
\affiliation{\FUW}
\author{Kenji~Watanabe}
\affiliation{\Watanabe}
\author{Takashi Taniguchi}
\affiliation{\Taniguchi}
\author{Adam Babi\'nski}
\affiliation{\FUW}
\author{Maciej R. Molas}
\email{maciej.molas@fuw.edu.pl}
\affiliation{\FUW}

\begin{abstract}
Optically dark excitonic states play a critical role in the valleytronic, electronic, and optical properties of monolayer semiconducting transition metal dichalcogenides. 
Here, we investigate how electrostatic doping affects the in-plane magnetic-field-induced activation of dark excitonic complexes in a gated WSe$_2$ monolayer. 
By continuously tuning the carrier density via gate voltage, we access $n$-type, charge-neutral, and $p$-type regimes and track the corresponding brightening dynamics. 
We find that the brightening rates of the dark negative trion ($T^{D-}$), dark neutral exciton ($X^{D}$), and dark positive trion ($T^{D+}$) exhibit a strong and nontrivial dependence on doping. 
In particular, the pronounced asymmetry in the brightening behaviour of the neutral $X^{D}$ complex and the charged $T^{D-}$ and $T^{D+}$ trions reveals distinct underlying carrier interactions, which we describe using a rate-equation model for their steady-state populations.
These findings highlight the key role of dark excitonic complexes in governing the optical response and carrier dynamics of doped S-TMD monolayers.
\end{abstract}

\maketitle

\begin{figure}[t]
		\subfloat{}%
		\centering
		\includegraphics[width=1\linewidth]{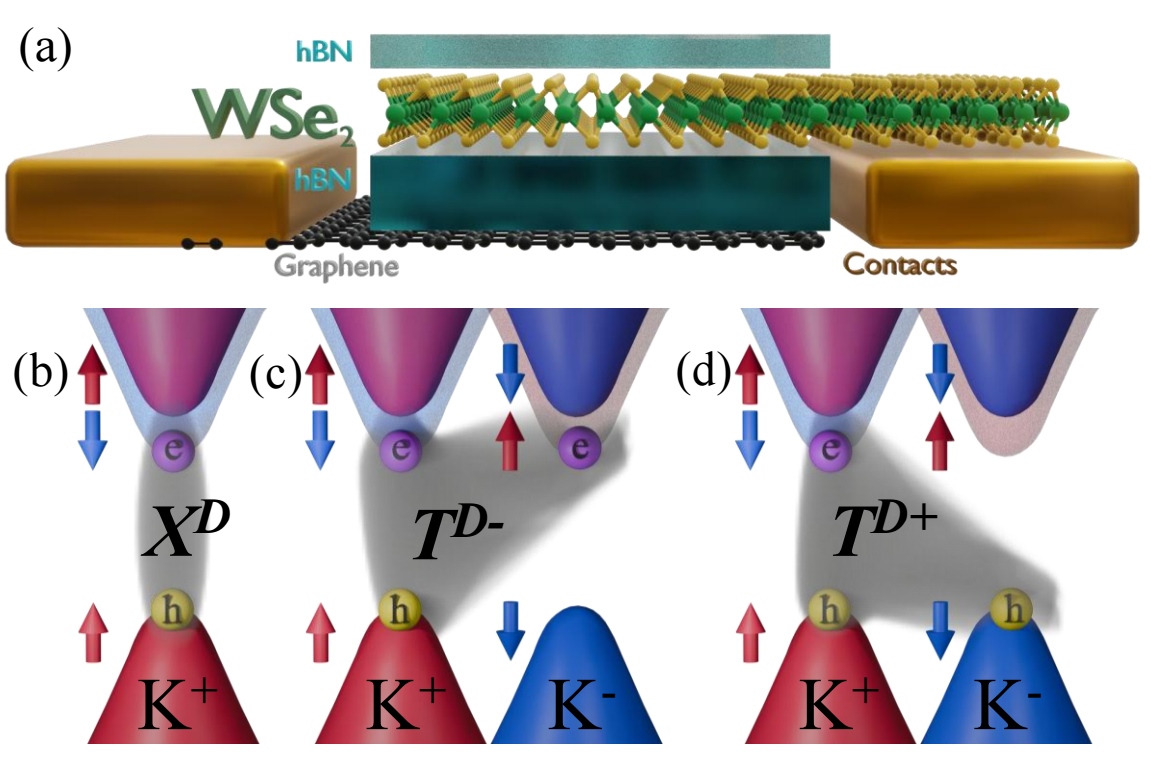}
        \caption{(a) Schematic illustration of the investigated heterostructure.
        The gated device consists of hBN/WSe$_2$ ML/hBN/graphene, with Pt contacts to the graphene layer and the WSe$_2$ ML. 
        Applying an external voltage enables control of the carrier density in the WSe$_2$ ML. 
        (b-d) Schematic diagrams of (b) the dark exciton ($X^{D}$), (c) the dark negative trion ($T^{D-}$), and (d) the dark positive trion ($T^{D+}$).}
		\label{fig:1_sample_structure}
\end{figure}

The optical response of monolayers (MLs) of semiconducting transition metal dichalcogenides (S-TMDs) is governed by a rich landscape of excitonic complexes and their dynamics, forming a versatile platform for exploring many-body interactions~\cite{Xu2014_TMDS_Spin,Mueller2018_TMDS_Excitons, Colloquim20218_TMDS_Excitons,  Mak2022_TMDS_Morie}. 
Among these systems, WSe$_2$ MLs have emerged as a prototypical model system, owing to their dark (optically inactive) excitonic ground state~\cite{Molas_2017_DX_BDX,Zhang2017_DX, Molas2019_DX, Kipczak2024_DX_BDX_WSe2}. 
Electrostatic gating of a WSe$_2$ ML, schematically illustrated in Fig.~\ref{fig:1_sample_structure}(a), enables continuous and reversible tuning of the carrier density, from the $n$-type regime, across the charge neutrality point, to $p$-type doping.
This tunability gives rise to a rich variety of excitonic complexes, clearly manifested in low-temperature ($T \sim 5$~K) photoluminescence (PL) spectra~\cite{Jones2013_WSE2_TuneNP, Molas_2017_DX_BDX, Robert2017_DX, Zhang2017_DX, Molas2019_DX, Liu2019_TuneNP_DX, Liu2019_TuneNP_DX_2, He2020_TuneNP_DX, Liu2020_WSE2_TuneNP, Zinkiewicz2020_DX_WSe2, Aurora2020_DX_WSe2, Zinkiewicz2021_DX,Kapuscinski2021_DX, Kipczak2024_DX_BDX_WSe2,  Jindal2025_WSE2_TuneNP}.
Despite this extensive tunability, a key aspect remains unresolved: the brightening dynamics of dark excitonic complexes in an in-plane magnetic field are still poorly understood, yet they are central to the emission properties of WSe$_2$ MLs.

The origin of this limitation lies in the intrinsic difficulty in accessing dark excitonic states.
Schematic representations of the energetically lowest configurations of $X^{D}$ and dark charged excitons ($T^{D-}$ and $T^{D+}$, respectively) are shown in Fig.~\ref{fig:1_sample_structure}(b-d). 
As a result of spin-conserving optical selection rules, their radiative recombination is nominally forbidden. 
Nevertheless, emission from grey excitons, which possess a finite out-of-plane dipole moment, as well as from momentum-forbidden dark trions enabled by disorder, has been observed at low temperatures even in the absence of external perturbations~\cite{Wang2017_WSe2, Molas2019_DX, Li2019_TuneNP_WSe2, Li2019_WSe2_TuneNP, Liu2020_WSE2_TuneNP, Zinkiewicz2021_DX, Pucko2023}.
A more direct route to activating these states is provided by external magnetic fields.
In particular, the radiative recombination of strictly spin-forbidden dark states can be activated via bright-dark exciton mixing induced by an in-plane magnetic field ($B$)~\cite{Slobodeniuk2016, Molas_2017_DX_BDX, Zhang2017_DX, Molas2019_DX}.
Consequently, the in-plane brightened emission intensity of dark excitonic complexes ($I_{D}$) is predicted to scale quadratically with $B$~\cite{Slobodeniuk2016, Molas_2017_DX_BDX, Zhang2017_DX, Molas2019_DX}, enabling direct observation of the $X^{D}$ exciton~\cite{Kapuscinski2021_DX}.
Although previous studies have provided important insights into the brightening of emission from dark neutral and charged excitons and their mutual interactions~\cite{Molas_2017_DX_BDX, Zhang2017_DX, Molas2019_DX, Zinkiewicz2020_DX_WSe2, Zinkiewicz2021_DX, Kipczak2024_DX_BDX_WSe2}, a comprehensive and systematic understanding of their behaviour across different doping regimes remains lacking.

Here, we investigate how electrostatic doping influences the activation of dark excitonic complexes by an in-plane magnetic field in a gated WSe$_2$ ML.
Low-temperature micro-PL measurements at $T=10$~K in in-plane magnetic fields up to 16~T provide direct insight into the brightening dynamics of dark states.
By continuously tuning the carrier density via gate voltage, we access the $n$-type, charge-neutral, and $p$-type regimes, and track the corresponding evolution of the emission spectra.
We find that the brightening rates of the $T^{D-}$, $X^{D}$, and $T^{D+}$ complexes exhibit a pronounced and nontrivial dependence on the doping level.
In particular, the marked asymmetry in the brightening of dark excitons and trions reveals distinct underlying carrier interactions, which we capture using a rate-equation model describing their steady-state populations. 
Overall, our results establish the central role of dark excitonic complexes in governing both the optical response and carrier dynamics in doped S-TMD MLs.

\section*{Results and discussion \label{sec:result}}
\subsection*{Doping-dependent brightening of dark excitons and trions}

\begin{figure}[t]
		\subfloat{}%
		\centering
		\includegraphics[width=1\linewidth]{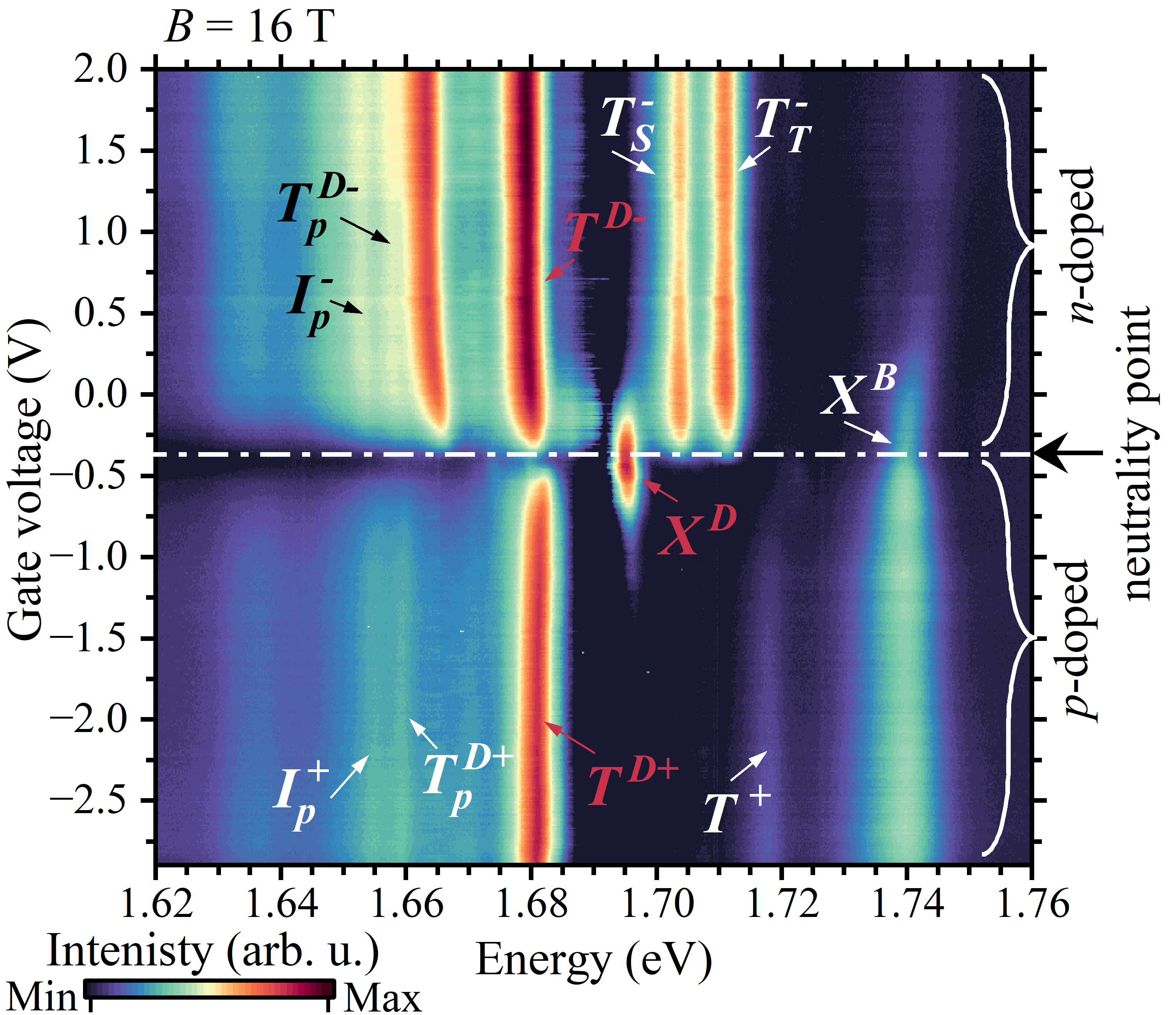}
        \caption{False-colour map of the low-temperature ($T=10$~K) PL of a WSe$_2$ ML versus gate voltage, measured in an in-plane magnetic field of 16~T. 
        The $n$-type, charge-neutral, and $p$-type regimes are indicated, with the neutrality point marked.
        The observed excitonic complexes, including bright ($X^{B}$, $T^{+}$, $T^{-}_{S}$, and $T^{-}_{T}$), dark ($X^{D}$, $T^{D-}$, and $T^{D+}$), and phonon replicas of dark complexes ($T^{D+}_{p}$, $T^{D-}_{p}$, $I^{+}_{p}$, and $I^{-}_{p}$), are labeled. Their origin is discussed in the text.}
		\label{fig:2_ESweeps}
\end{figure}

The influence of doping on the in-plane brightened PL emission of dark excitonic complexes in a WSe$_2$ ML is investigated using the architecture shown in Fig.~\ref{fig:1_sample_structure}. 
Gate-dependent PL spectra measured in an in-plane magnetic field of 16~T are presented as a false-colour map in Fig.~\ref{fig:2_ESweeps}, with the corresponding zero-field data shown in Fig.~S1 in the Supplementary Information (SI).
Three distinct doping regimes are identified: $n$-type (from 2 to $-0.35$~V), where emission from negatively charged complexes is predominant ($T^{-}_{S}$, $T^{-}_{T}$, $T^{D-}$, $T^{D-}_{p}$, and $I^{-}_{p}$); charge-neutral (from $-0.35$ to $-0.55$~V), where charged-complex emission is suppressed and neutral excitons dominate ($X^{B}$ and $X^{D}$); and $p$-type (from $-0.55$ to $-3$~V), where positively charged complexes prevail ($T^{+}$, $T^{D+}$, $T^{D+}_{p}$, and $I^{+}_{p}$).
In addition to the main bright excitonic complexes present across all regimes, namely neutral exciton ($X^{B}$), intravalley spin-singlet ($T^{-}_{S}$) and intervalley spin-triplet ($T^{-}_{T}$) negative trions, and intervalley spin-triplet ($T^{+}$) positive trion, we resolve three distinct emission lines attributed to their dark counterparts ($X^{D}$, $T^{D-}$, and $T^{D+}$), along with a series of phonon replicas ($T^{D+}_{p}$, $T^{D-}_{p}$, $I^{+}_{p}$, and $I^{-}_{p}$).
All spectral features are identified based on previous reports~\cite{Jones2013_WSE2_TuneNP, Molas_2017_DX_BDX, Robert2017_DX, Zhang2017_DX, Courtade2017_WSe_TuneNP, Chen2018_WSe, Molas2019_DX, Li2019_WSe2_TuneNP, Liu2019_TuneNP_DX, Liu2019_TuneNP_DX_2, Liu2020_WSE2_TuneNP, Aurora2020_DX_WSe2, He2020_TuneNP_DX, Kipczak2024_DX_BDX_WSe2, Jindal2025_WSE2_TuneNP}.

Using a simple parallel-plate capacitor model, we estimate the free carrier concentration at selected gate voltages (see Section~S2 in the SI for details).
In particular, we assume that a voltage of $-0.45$~V corresponds to the charge neutrality point, while increasing or decreasing the gate voltage results in a linear change in the doping level.
This yields free electron and hole concentrations on the order of $1 \times 10^{12}$~cm$^{-2}$ for gate voltages of 1~V and $-2$~V, respectively.

The neutrality point regime, corresponding to a negligible concentration of free carriers, reveals a significantly different spectrum compared to that in the doped regimes. 
While the $X^{B}$ line persists across the entire range of applied voltages, displaying a characteristic blueshift with increasing $n$-type doping, the $X^{D}$ emission remains strongly confined to a narrow voltage window around the neutrality point.
A similar trend has been reported for the neutral grey exciton ($X^{G}$), a state characterised by an out-of-plane dipole moment, which can be detected even in the absence of an external magnetic field due to the high numerical aperture of the microscope objectives used in PL experiments (see Fig.~S1 in the SI)~\cite{Wang2017_WSe2, Robert2017_DX, Molas2019_DX, Zinkiewicz2020_DX_WSe2, Kipczak2024_DX_BDX_WSe2, Li2019_WSe2_TuneNP, Liu2019_TuneNP_DX, Liu2019_TuneNP_DX_2, Liu2020_WSE2_TuneNP, He2020_TuneNP_DX, Jindal2025_WSE2_TuneNP}.
In contrast, the appearance of the $X^{D}$ line originates exclusively from the applied in-plane $B$ field, as this state possesses a vanishing dipole moment~\cite{Robert2017_DX, Molas2019_DX, Zinkiewicz2020_DX_WSe2, Kipczak2024_DX_BDX_WSe2}.
Furthermore, the intensity of the $X^{D}$ line is substantially enhanced as a result of the magnetic-field-induced brightening mechanism, especially compared to that of $X^{B}$~\cite{Molas_2017_DX_BDX, Zhang2017_DX, Zinkiewicz2020_DX_WSe2, Kapuscinski2021_DX, Kipczak2024_DX_BDX_WSe2}.

\begin{figure*}[t!]
		\subfloat{}%
		\centering
		\includegraphics[width=1\linewidth]{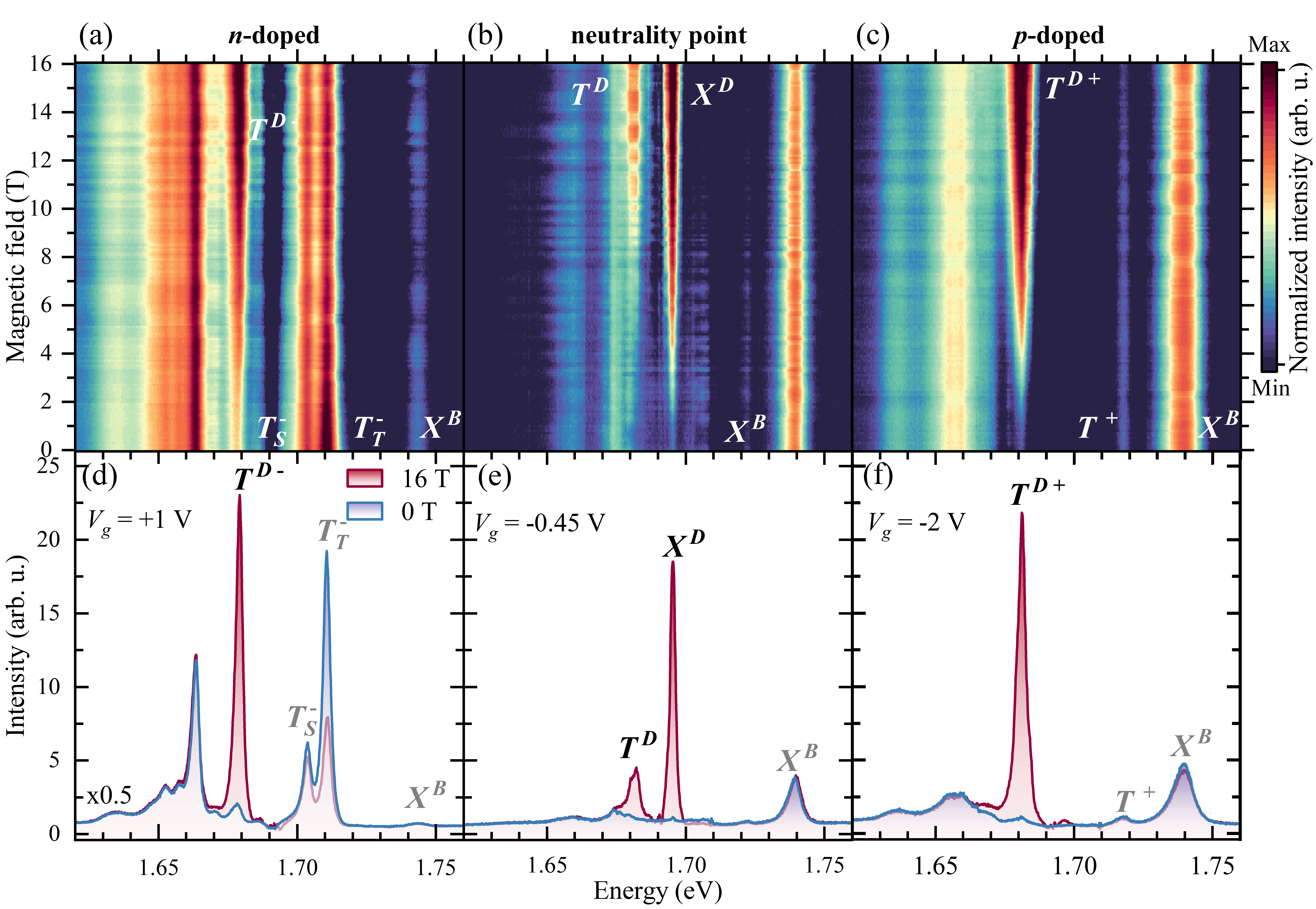}
        \caption{(a)-(c) False-colour maps of low-temperature ($T=10$~K) PL of a WSe$_2$ ML versus in-plane magnetic field for three representative gate voltages: (a) $n$-doped regime, (b) neutrality point, and (c) $p$-doped regime. 
        The colour-coded intensities in panels (a)-(c) are normalised to the maximum intensity in each panel for clarity.
        (d)-(f) Comparison of the corresponding PL spectra in the three regimes measured at zero magnetic field (blue curves) and at $B=16$~T (red curves).
        The intensities in panels (d)–(f) correspond to the absolute values measured in the experiment. 
        The intensity in panel (d) is therefore scaled by a factor of 0.5.}
		\label{fig:3_Maps}
\end{figure*}

Increasing or decreasing the gate voltage beyond the neutrality point leads to the emergence of free electrons or holes, respectively, corresponding to $n$- and $p$-type doping of the studied WSe$_2$ ML.
Owing to the asymmetry between the conduction and valence bands of the ML (see Fig.~\ref{fig:1_sample_structure}), two distinct lines, $T^{-}_{S}$ and $T^{-}_{T}$, are attributed to negative trions, whereas a single $T^{+}$ feature corresponds to the positive trion, in agreement with previous reports~\cite{Courtade2017_WSe_TuneNP, Liu2020_WSE2_TuneNP, He2020_TuneNP_DX}.
In contrast, dark trions exhibit a single spectral feature within a given doping regime ($T^{D-}$ and $T^{D+}$), reflecting their analogous carrier configurations~\cite{Liu2019_TuneNP_DX, Li2019_TuneNP_WSe2, Li2019_WSe2_TuneNP, Liu2019_TuneNP_DX_2}.
Notably, the fine structure of dark trions differs from that of neutral dark excitons ($X^{G}$ and $X^{D}$).
However, studies of WS$_2$ MLs indicate that these transitions possess an intervalley character at zero magnetic field, whereas intravalley recombination channels become active in the presence of a finite in-plane magnetic field~\cite{Zinkiewicz2021_DX}.
Due to the uncertainty in assigning the charge sign to dark trions near the neutrality point, this feature is denoted as $T^{D}$.
On the low-energy side in Fig.~\ref{fig:2_ESweeps}, a series of phonon replicas of dark trions ($T^{D+}_{p}$, $T^{D-}_{p}$, $I^{+}_{p}$, and $I^{-}_{p}$) can be resolved~\cite{Liu2020_WSE2_TuneNP,He2020_TuneNP_DX, Robert2021_TuneNP_WSe2,Jindal2025_WSE2_TuneNP}, see the SI for details. 
The evolution of excitonic features as a function of gate voltage at 16~T in both doping regimes closely follows that observed at zero magnetic field (see the Fig.~S1) and remains consistent with previous studies~\cite{Courtade2017_WSe_TuneNP, Li2018_Tune_WSe2, Barbone2018_TuneNP_WSe2, Liu2019_TuneNP_DX, Li2019_TuneNP_WSe2, Li2019_WSe2_TuneNP,Liu2019_TuneNP_DX_2, Liu2020_WSE2_TuneNP,He2020_TuneNP_DX, Robert2021_TuneNP_WSe2,Jindal2025_WSE2_TuneNP}.

To verify the $B$-field evolution of dark excitons and trions, Fig.~\ref{fig:3_Maps}(a)-(c) presents false-colour maps of the low-temperature ($T=10$~K) PL emission of a WSe$_2$ ML versus in-plane magnetic field for three representative gate voltages: the $n$-doped regime, the neutrality point, and the $p$-doped regime.
While the intensities of most emission lines, in particular $X^{B}$, remain nearly unchanged, with only minor variations that can be attributed to slight experimental instability, the intensities of the dark complexes increase markedly with magnetic field.
At low magnetic fields, the intensities of the $X^{D}$, $T^{D}$, $T^{D-}$, and $T^{D+}$ features are comparable to those of other emission lines.
With increasing $B$, these features become progressively more pronounced and dominate the spectra at 16~T as a result of the efficient brightening effect.

Figures~\ref{fig:3_Maps}(d)–(f) present a comparison of the corresponding PL spectra in the three doping regimes measured at zero magnetic field and at $B=16$~T.
These data confirm that the applied in-plane $B$ field does not significantly affect most optically active transitions, such as $X^{B}$ and $T^{+}$.
However, a pronounced reduction in the intensities of the $T^{-}_{S}$ and $T^{-}_{T}$ features is observed, which may be associated with the brightening of the $T^{D-}$ trion, as discussed below.
In all investigated doping regimes, the brightened $X^{D}$, $T^{D}$, $T^{D-}$, and $T^{D+}$ emissions are clearly resolved.
Moreover, at 16~T, the $T^{D-}$ feature exhibits the highest intensity (the PL spectrum in Fig.~\ref{fig:3_Maps}(d) is scaled by a factor of 0.5 for clarity), followed by $T^{D+}$ in Fig.~\ref{fig:3_Maps}(f), while $X^{D}$ remains the weakest in Fig.~\ref{fig:3_Maps}(e).
This hierarchy indicates that the in-plane $B$-field-induced brightening effect depends strongly on the doping regime.

\begin{figure}[t]
		\subfloat{}%
		\centering
		\includegraphics[width=1\linewidth]{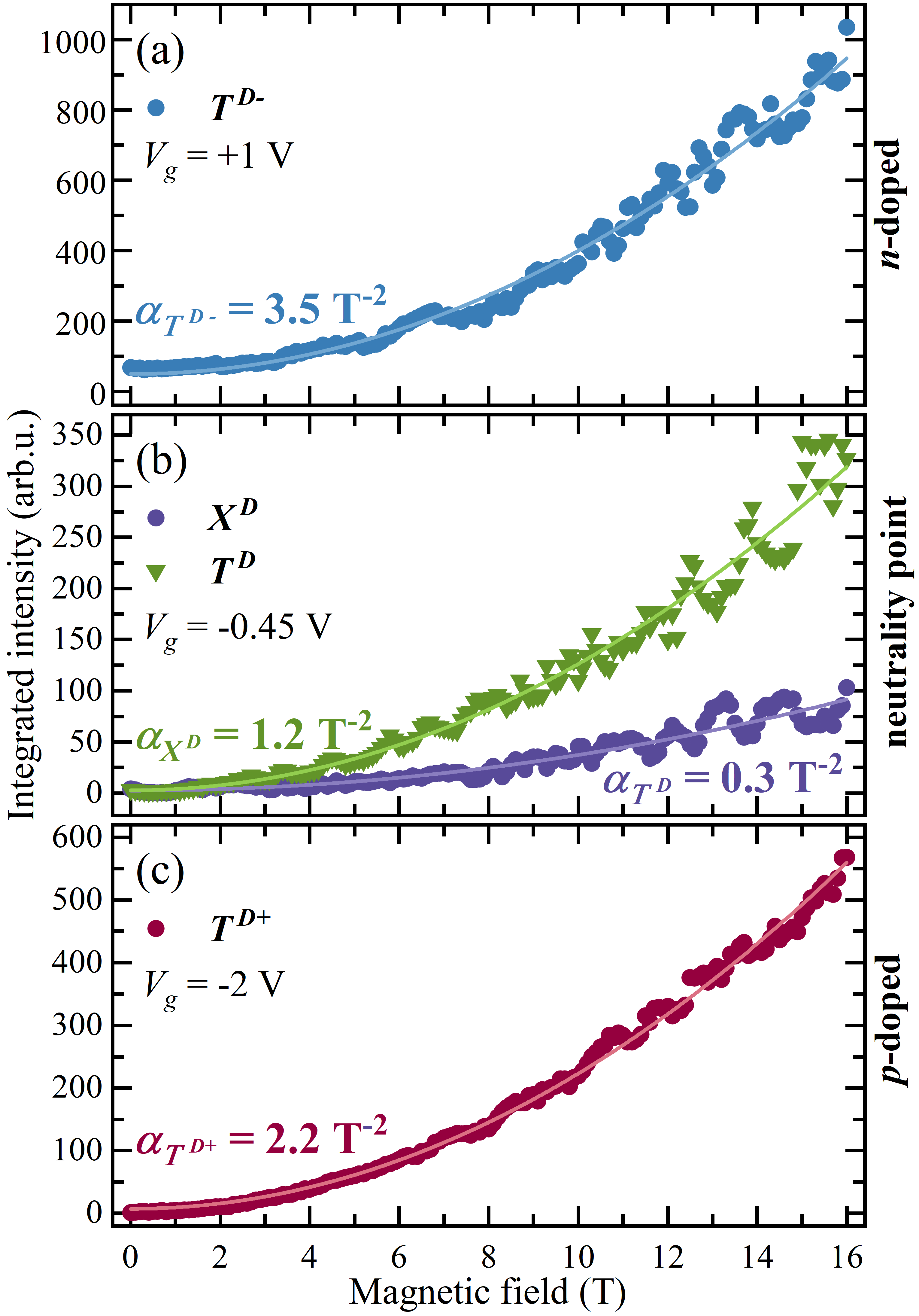}
        \caption{Magnetic-field dependence of the integrated intensities of dark excitonic complexes extracted for three representative gate voltages: (a) $n$-doped regime, (b) the neutrality point, and (c) $p$-doped regime.
        Coloured curves represent fits to the function $I_D = I_0 + \alpha B^2$.}
		\label{fig:4}
\end{figure}

To analyse in detail the doping effect on the brightening of dark complexes, the PL spectra were deconvoluted using Lorentzian functions. 
The extracted evolution of the integrated intensities ($I_D$) of the $X^{D}$, $T^{D}$, $T^{D-}$, and $T^{D+}$ complexes as a function of the in-plane magnetic field for three selected gate voltages is presented in Fig.~\ref{fig:4}. 
These dependencies are expected to follow a quadratic dependence and can be described by $I_D = I_0 + \alpha B^2$, where $I_0$ is a fitting parameter corresponding to the zero-field intensity, and $\alpha$ denotes the brightening factor~\cite{Slobodeniuk2016, Molas_2017_DX_BDX, Zhang2017_DX, Molas2019_DX, Kipczak2024_DX_BDX_WSe2}. 
The parameter $\alpha$ is proportional to the population of dark complexes and to the emission intensity of their bright analogues~\cite{Slobodeniuk2016, Molas_2017_DX_BDX}. 
As shown in Fig.~\ref{fig:4}, the experimental data are well reproduced by the fitted curves. 
From these fits, the brightening factors are extracted: $\alpha_{T^{D-}} = 3.5$~T$^{-2}$ for the $n$-doped regime, $\alpha_{X^{D}} = 1.2$~T$^{-2}$ for the neutrality point, and $\alpha_{T^{D+}} = 2.2$~T$^{-2}$ for the $p$-doped regime. 
These values reveal pronounced differences between the complexes, with the brightening factor of $X^{D}$ being significantly smaller than those of both dark trions. The brightening factors of $T^{D-}$ and $T^{D+}$ also differ substantially from each other.

\begin{figure}[t]
		\subfloat{}%
		\centering
		\includegraphics[width=1\linewidth]{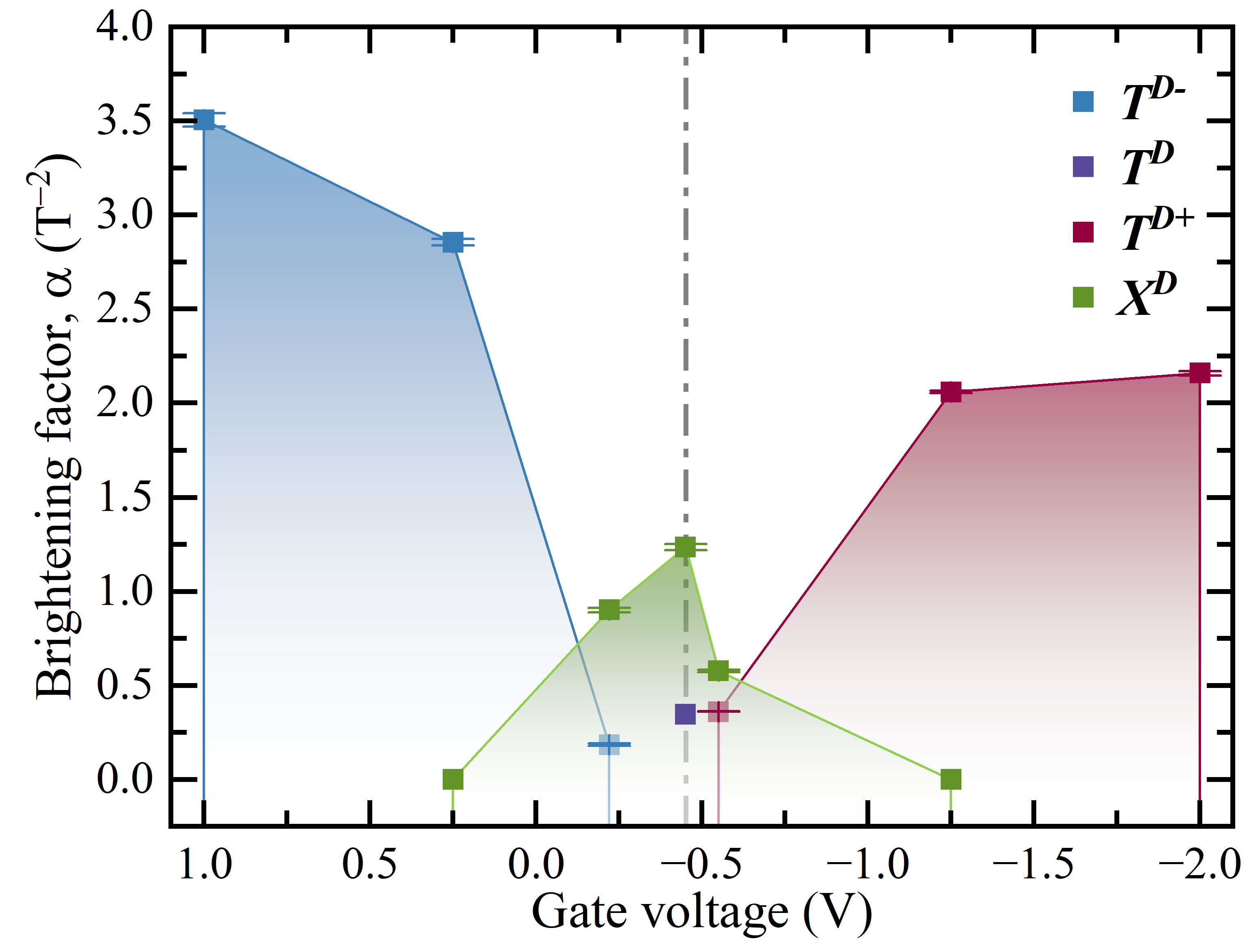}
        \caption{Extracted brightening factors ($\alpha$) of the $X^{D}$, $T^{D}$, $T^{D-}$, and $T^{D+}$ transitions as a function of the gate voltage.
        The vertical dashed line marks the charge neutrality point.}
		\label{fig:5}
\end{figure}

Further insight into this brightening behaviour is obtained from a systematic series of magnetic-field-dependent PL measurements performed at different gate voltages: $-1.25$~V, $-0.55$~V, 0.22~V, and 1.25~V.
False-colour magneto-PL maps, together with the corresponding integrated intensities and quadratic fits, are presented in Section S3 of the SI.
The extracted brightening factors for the different dark complexes as a function of gate voltage are summarised in Fig.~\ref{fig:5}.
The brightening factor of the $X^{D}$ exciton reaches a maximum value of approximately 1.2~T$^{-2}$ at the neutrality point and is rapidly quenched in both the $n$- and $p$-doping regimes.
The $\alpha$ parameters of the $T^{D+}$ and $T^{D-}$ trions increase significantly with increasing carrier density, reaching values of 2.2~T$^{-2}$ and 3.5~T$^{-2}$, respectively, at the highest carrier concentrations.
These results reveal a pronounced asymmetry in the brightening effects not only between neutral and charged dark complexes but also between the dark trions themselves, pointing to distinct underlying carrier dynamics. The theoretical model presented below captures this phenomenon well.

\subsection*{Rate-equation framework for dark excitonic complexes}

To rationalise the markedly different brightening efficiencies of the $T^{D-}$ and $T^{D+}$ dark trions relative to the $X^{D}$ dark exciton, we base our analysis on three key experimental observations.
First, the dark trion energies, $E(T^{D-}) = 1.679$~eV and $E(T^{D+}) = 1.681$~eV, are significantly lower than that of the neutral dark exciton, $E(X^{D}) = 1.696$~eV. 
Consequently, thermal occupation of dark trion states is favoured over that of $X^{D}$.
Second, under sufficiently large negative (positive) gate voltages, the $X^{D}$ emission is completely quenched (see Fig.~\ref{fig:2_ESweeps}), indicating efficient conversion of $X^{D}$ into charged complexes once the electron or hole density exceeds a threshold value.
Third, $X^{B}$ emission remains strong in the $p$-doped regime, whereas its intensity decreases significantly under $n$-doping. 
This asymmetry indicates that distinct processes govern the dynamics of exciton and trion populations in S-TMD monolayers under $p$- and $n$-doping conditions~\cite{Liu2019_TuneNP_DX, Li2019_WSe2_TuneNP, Courtade2017_WSe_TuneNP}.

Since the PL intensity of each spectral feature is proportional to the population of the corresponding excitonic complex, a quantitative interpretation of the intensity evolution requires estimating these populations as functions of carrier density and temperature.
We capture this behaviour using a simple rate-equation model, treating electron and hole doping separately.
This model provides a qualitative, physically motivated framework for interpreting the observed trends; however, a quantitative description of the experimental data requires a more sophisticated extension of the model, as discussed, for example, in Ref.~\cite{Baranowski2017}.

We first consider the electron-doped regime.
Let $N$ and $N^{-}$ denote the populations of the $X^{D}$ and $T^{D-}$ complexes, respectively, within a given valley ($\mathrm{K}^{+}$ or $\mathrm{K}^{-}$) of an S-TMD ML.
Application of a positive gate voltage populates the lower conduction band (see Fig.~\ref{fig:1_sample_structure}(b)), resulting in an electron density $n$ that increases monotonically with the applied voltage.
Experimentally, this regime is characterised by a pronounced enhancement of the $T^{D-}$ emission, accompanied by a simultaneous suppression of the $X^{D}$ line, as shown in Fig.~\ref{fig:2_ESweeps}.
 
We assume that $T^{D-}$ trions form via the binding of $X^{D}$ with free electrons.
The $X^{D}$ states are generated from optically bright excitons through various mechanisms (see, for example, Refs.~\cite{Yang2022, Mai2014, Sie2015} and references therein).
$X^{B}$ formed via optical transitions at the band extrema, remain efficiently generated under electron doping. 
However, owing to their higher energy, they relax into lower-energy excitonic complexes, including $T^{-}_{T}$, $T^{-}_{S}$, $X^{D}$ and $T^{D-}$.
The efficient formation of these complexes depletes the bright exciton line, as observed in Figs.~\ref{fig:2_ESweeps} and \ref{fig:3_Maps}(a) and (d).

The aforementioned population dynamics of the exciton complexes $T^{D-}$ and $X^D$ are described by the following rate equations.
\begin{align}
\frac{dN}{dt}=&\, g -\frac{N}{\tau_\text{d}} - \gamma_- N n + \frac{N^-}{\tau_-} e^{-\frac{\Delta E^-}{kT}}, \\
\frac{dN^-}{dt}=&\, -\frac{N^-}{\tau_\text{t-}} + \gamma_- N n - \frac{N^-}{\tau_-} e^{-\frac{\Delta E^-}{kT}},
\end{align}
where $g$ denotes the generation rate of $X^D$, $\tau_\text{d}$ and $\tau_\text{t-}$ are the effective lifetimes of $X^D$ and $T^{D-}$, respectively, and $\gamma_-$ describes the efficiency of trion formation via exciton-electron capture. 
The reverse process, namely the thermal dissociation of a trion into a neutral dark exciton and a free electron, is characterised by the timescale $\tau_-$ and is suppressed by a Boltzmann factor associated with the energy difference $\Delta E^- = E(X^D)-E(T^{D-})$.

In steady state, the populations are given by
\begin{align}
N^-=&\, g \tau_\text{t-}
\frac{\gamma_- n \tau_\text{d}}{1 + \gamma_- n \tau_\text{d} + \frac{\tau_\text{t-}}{\tau_-} e^{-\frac{\Delta E^-}{kT}}}, \\
N =&\, g \tau_\text{d}
\frac{1 + \frac{\tau_\text{t-}}{\tau_-} e^{-\frac{\Delta E^-}{kT}}}{1 + \gamma_- n \tau_\text{d} + \frac{\tau_\text{t-}}{\tau_-} e^{-\frac{\Delta E^-}{kT}}}.
\end{align}
For $\Delta E^- \gg kT$, relevant to our experiments, trion dissociation is negligible and the trion population simplifies to
\begin{align}
N^- \approx N \gamma_- n \tau_\text{t-}.
\end{align}
Thus, once $\gamma_- n \tau_\text{t-} \gtrsim 1$, $T^{D-}$ trions dominate the excitonic population. 
This explains the experimentally observed crossover, where the $T^{D-}$ emission becomes stronger than the $X^{D}$ line at sufficiently high electron densities.

We now turn to the hole-doped regime, realised by applying a negative gate voltage.
In this case, the top of the valence band is progressively depleted, leading to a hole concentration $p$ in the ML.
The populations of $X^{D}$ and $T^{D+}$, denoted by $N$ and $N^{+}$, respectively, are described by the following set of equations:
\begin{align}
\frac{dN}{dt}=&\, \widetilde{g} -\frac{N}{\tau_\text{d}} - \gamma_+ N p + \frac{N^+}{\tau_+} e^{-\frac{\Delta E^+}{kT}}, \\
\frac{dN^+}{dt}=&\, -\frac{N^+}{\tau_\text{t+}} + \gamma_+ N p - \frac{N^+}{\tau_+} e^{-\frac{\Delta E^+}{kT}},
\end{align}
where $\widetilde{g}$ denotes the generation rate of $X^D$, $\tau_\text{d}$ and $\tau_\text{t+}$ are the effective lifetimes of $X^D$ and $T^{D+}$, respectively, and $\gamma_+$ describes the efficiency of trion formation via exciton-hole capture. 
As in the case of $T^{D-}$, the reverse process, $i.e.$, thermal dissociation of a trion into $X^D$ and a hole, is characterised by the timescale $\tau_+$ and is suppressed by a Boltzmann factor associated with the energy difference $\Delta E^+ = E(X^D)-E(T^{D+})$.

Although the structure of these equations is analogous to that in the electron-doped case, the key difference lies in the generation rate $\widetilde{g}$ of $X^D$. 
Compared to the $n$-doped regime, the bright exciton line is not suppressed under $p$-doping, as shown in Figs.~\ref{fig:3_Maps}(c) and (f), reflecting a significantly slower conversion of bright excitons into lower-energy excitonic complexes~\cite{Yang2022}.
Consequently, only a smaller fraction of the bright exciton population participates in the formation of dark excitons, implying that the generation rate $\widetilde{g}$ in the $p$-doped regime is lower than $g$ in the $n$-doped regime.

As in the previous case, the steady-state population of $T^{D+}$ dominates over that of dark excitons ($N$):
\begin{align}
N^+ \approx N \gamma_+ p \tau_\mathrm{t_+},
\end{align}
for relatively large $\gamma_+ p \tau_\mathrm{t_+} \gtrsim 1$. 
However, in the $p$-doped regime, the total number of dark excitons ($N \propto \widetilde{g}$) is smaller than in the $n$-doped regime ($N \propto g$), reflecting the fact that $\widetilde{g} < g$, as discussed above.
Consequently, under identical experimental conditions, the PL intensity of brightened $T^{D+}$ in the $p$-doped regime is lower than that of $T^{D-}$ in the $n$-doped regime.

\section{Summary \label{sec:Summary}}
In summary, we examine the influence of electrostatic doping on the in-plane magnetic-field-induced activation of dark excitonic complexes in a gated WSe$_2$ ML. 
By varying the gate voltage, we access electron-doped, charge-neutral, and hole-doped regimes and resolved the corresponding evolution of the emission spectra. 
We find that the brightening efficiencies of $T^{D-}$, $X^{D}$, and $T^{D+}$ depend strongly and nonlinearly on carrier density. 
Notably, the contrasting behaviour of neutral and charged dark complexes points to distinct microscopic pathways governing their formation and recombination, which we capture using a steady-state rate-equation model. 
These results establish electrostatic doping as an effective means of controlling dark exciton populations and the resulting optical response in S-TMD monolayers.

\section*{Methods \label{sec:methods}}
\subsection*{Sample fabrication}
The investigated sample consists of an hBN capping flake, a WSe$_2$ ML, an hBN insulating barrier, and a graphite flake, as shown in Fig.~\ref{fig:1_sample_structure}. 
The WSe$_2$ ML and hBN flakes were mechanically exfoliated using dicing tape onto polydimethylsiloxane (PDMS) and subsequently transferred onto a 295~nm SiO$_2$/Si substrate. 
Graphite flakes were exfoliated directly onto the same type of substrate. 
The heterostructure was assembled using a deterministic dry-transfer technique with a polycarbonate (PC) film supported on a PDMS stamp at 90~$^\circ$C. 
The complete stack was then transferred onto a 295~nm SiO$_2$/Si substrate with pre-patterned Pt/Cr contacts at 180~$^\circ$C. 
Finally, the residual PC film was dissolved in chloroform.

\subsection*{Photoluminescence spectroscopy}
Low-temperature micro-magneto-PL experiments were performed in the Voigt geometry, $i.e.$, with the magnetic field oriented parallel to the ML plane. 
Measurements with a spatial resolution of $\sim 1~\mu$m were carried out using a superconducting magnet providing fields up to 16~T in a free-space optical configuration.
The sample was mounted on an $x$–$y$–$z$ piezoelectric stage maintained at $T=10$~K and excited by a continuous-wave laser diode with a wavelength of 515~nm (photon energy of 2.41~eV). 
The emitted light was dispersed by a spectrometer with a focal length of 0.5~m and detected using a CCD camera.

\section*{Data availability}
The datasets generated and analysed during the current study are publicly available at the following link: XXX.

\section{Acknowledgements}
The work has been supported by the National Science Centre, Poland (2022/46/E/ST3/00166).
K. W. and T. T. acknowledge support from the JSPS KAKENHI (Grants No. 21H05233 and No. 23H02052), the CREST (No. PMJCR24A5), JST and World Premier International Research Center Initiative (WPI), MEXT, Japan. 

\section*{Author contributions}
M.R.M. conceived the project, designed the experiments, and secured funding.
G.K. fabricated the sample.
G.K., K.W., K.O., and M.R.M. performed the optical measurements.
G.K. and M.R.M. carried out the data analysis.
A.O.S. conducted the theoretical studies.
G.K., A.B., and M.R.M. interpreted the experimental results.
T.T. and K.W. provided the hBN crystals.
G.K., A.O.S. and M.R.M. wrote the manuscript.
All authors contributed to the discussions and provided feedback on the manuscript.

\bibliographystyle{apsrev4-2}
\bibliography{biblio}

\newpage
\onecolumngrid

\renewcommand{\thefigure}{S\arabic{figure}}
\renewcommand{\thesubsection}{S\arabic{subsection}}
\renewcommand{\thetable}{S\Roman{table}}
\renewcommand{\theequation}{S\arabic{equation}}

\setcounter{section}{0}
\setcounter{figure}{0}
\setcounter{table}{0}

\begin{center}
	{\large{ {\bf Supplementary Information: \\ Doping-Induced Brightening of Dark Excitons and Trions in a WSe$_2$ Monolayer}}}
	\vskip0.5\baselineskip{Grzegorz Krasucki,{$^{1}$} Artur O. Slobodeniuk,{$^{2}$} Kacper Walczyk,{$^{1}$} Katarzyna Olkowska-Pucko,{$^{1}$} Kenji Watanabe,{$^{3}$} Takashi Taniguchi,{$^{4}$} Adam Babiński,{$^{1}$}  and Maciej R. Molas{$^{1}$}}
	
	\vskip0.5\baselineskip{\em $^{1}$ University of Warsaw, Faculty of Physics, 02-093 Warsaw, Poland \\$^{2}$ Charles University, Faculty of Mathematics and Physics,  CZ-121 16 Prague, Czech Republic 
		\\$^{3}$ Research Center for Electronic and Optical Materials, National Institute for Materials Science, 1-1 Namiki, Tsukuba 305-0044, Japan \\$^{4}$ Research Center for Materials Nanoarchitectonics, National Institute for Materials Science, 1-1 Namiki, Tsukuba 305-0044, Japan}
\end{center}

\subsection{Gate-dependent PL emission of WSe$_2$ monolayer}

\begin{figure}[h]
		\subfloat{}%
		\centering
		\includegraphics[width=86.36mm]{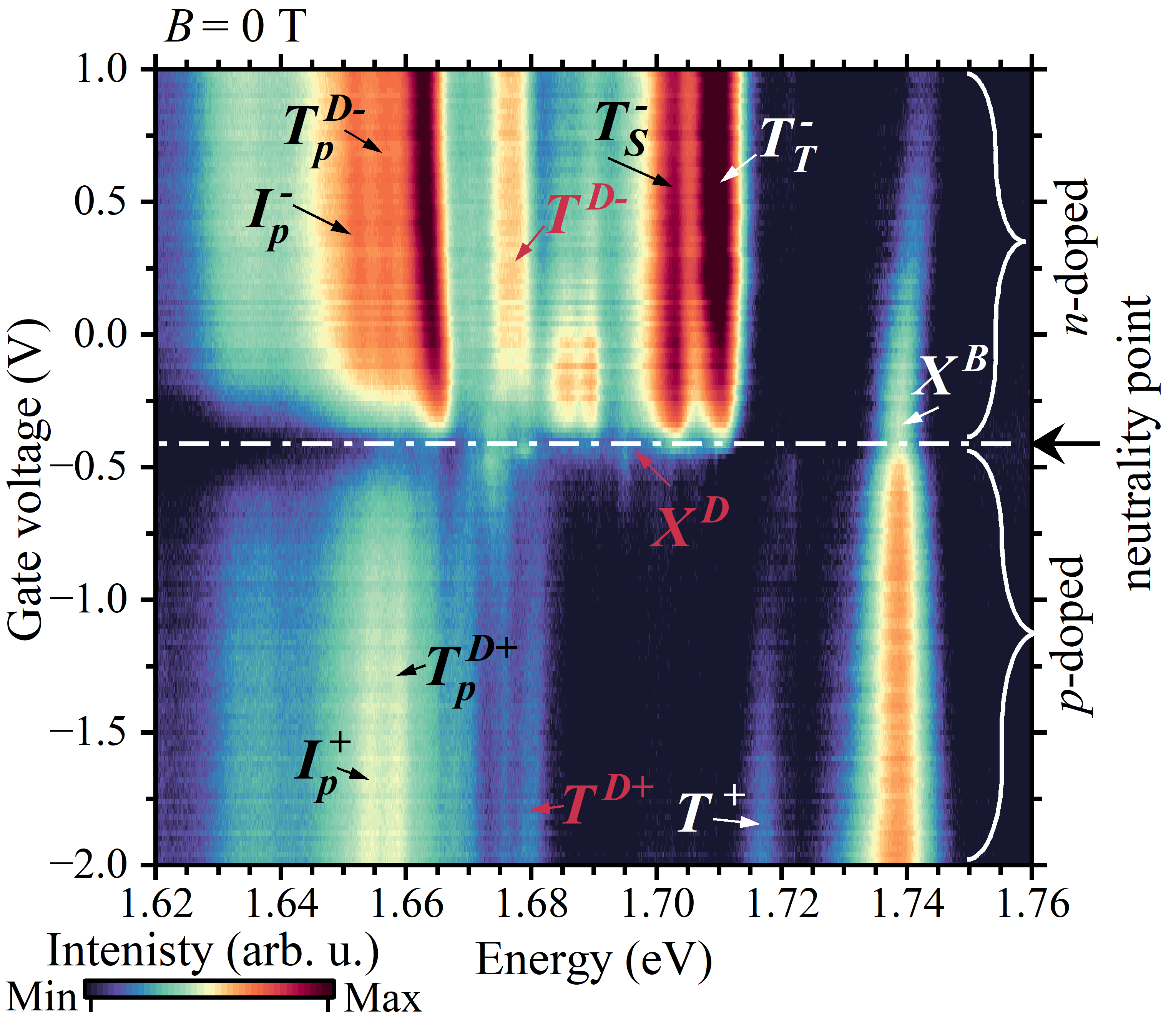}
        \caption{False-colour map of the low-temperature ($T=10$~K) PL of a WSe$_2$ ML versus gate voltage, measured at 0~T. 
        The $n$-type, charge-neutral, and $p$-type regimes are indicated, with the neutrality point marked.
        The various excitonic complexes, including bright ($X^{B}$, $T^{+}$, $T^{-}_{S}$, and $T^{-}_{T}$), dark ($X^{D}$, $T^{D-}$, and $T^{D+}$), and phonon replicas of dark complexes ($T^{D+}_{p}$, $T^{D-}_{p}$, $I^{+}_{p}$, and $I^{-}_{p}$), are labeled, and their origin is discussed in the text.}
		\label{fig:S1_0T}
\end{figure}

Figure~\ref{fig:S1_0T} shows a gate-dependent low-temperature ($T=10$~K) photoluminescence (PL) false-colour map of a monolayer (ML) WSe$_2$ at zero magnetic field. 
The PL spectrum exhibits a series of bright excitonic complexes, including neutral excitons ($X^{B}$), intravalley spin-singlet ($T^{-}_{S}$) and intervalley spin-triplet ($T^{-}_{T}$) negative trions, as well as the intervalley spin-triplet ($T^{+}$) positive trion. 
In the intermediate energy range, emission from the grey exciton ($X^{G}$) is observed together with its $n$- and $p$-doped counterparts, $i.e.$, negative and positive dark trions ($T^{D-}$ and $T^{D+}$), while at lower energies a series of phonon replicas ($T^{D+}_{p}$, $T^{D-}_{p}$, $I^{+}_{p}$, and $I^{-}_{p}$) can be resolved. 
The $T^{D+}_{p}$ and $T^{D-}_{p}$ lines correspond to intravalley optical transitions, $i.e.$, phonon replicas of dark trions arising from phonon emission at the $\Gamma$ point of the Brillouin zone (BZ), whereas the $I^{+}_{p}$ and $I^{-}_{p}$ lines originate from intervalley recombination processes enabled by phonon emission from the K valleys of the BZ. 
The assignment of all observed emission lines is supported by previous reports on gate-dependent PL spectra of ML WSe$_2$.~\cite{Jones2013_WSE2_TuneNP, Molas_2017_DX_BDX, Robert2017_DX, Zhang2017_DX, Courtade2017_WSe_TuneNP, Chen2018_WSe, Molas2019_DX, Li2019_WSe2_TuneNP, Liu2019_TuneNP_DX, Liu2019_TuneNP_DX_2, Liu2020_WSE2_TuneNP, Aurora2020_DX_WSe2, He2020_TuneNP_DX, Kipczak2024_DX_BDX_WSe2, Jindal2025_WSE2_TuneNP}.

\newpage

\subsection{Estimation of carrier concentration}

The carrier concentration in the gated device is estimated using a simple parallel-plate capasitor model.
By considering the applied voltage ($V$), the thickness of the dielectric barrier (hBN) $d$, and the dielectric constant of hBN, $\varepsilon_{hBN} \sim 3.7$~\cite{Ohba2001_ehBN,Laturia2020_ehBN}, the free electron density ($n$) can be expressed as $n = \frac{\varepsilon_0 \varepsilon_{hBN}}{e d}V$.
Owing to the presence of intrinsic doping in the WSe$_2$ ML, this relation enables the determination of the change in carrier concentration as a function of gate voltage.
As shown in Fig.~2 of the main article and in Fig.~\ref{fig:S1_0T}, the charge neutrality point, corresponding to zero carrier concentration, is located at approximately 0.45~V.
Accordingly, the above expression can be rewritten as follows:
\begin{equation}
n=\frac{\varepsilon_0 \varepsilon{hBN}}{ed}(V_g+0.45~\text{V})
\end{equation}
The thickness of the hBN barrier in the investigated device was determined to be 29~nm using atomic force microscopy.
Based on this value and the relation above, the carrier concentration at the selected gate voltages was estimated.
The resulting values are summarised in Table~\ref{tab:S2}.
The obtained carrier densities are consistent with those reported in the literature for similar devices with comparable hBN barrier thicknesses~\cite{Robert2021_TuneNP_WSe2}.

\begin{table}[th]
\centering
\caption{Estimated carrier concentration ($n$) for the gate voltages used in the article. }
\begin{tabular}{|c|c|c|}
\hline
$V_g$ (V) & $n$ (cm$^{-2}$) \\
\hline
+1.00	&	-1.04 $10^{12}$	\\
+0.25	&	-5.02 $10^{11}$	\\
-0.22	&	-1.65 $10^{11}$	\\
-0.45	&	neutrality point	\\
-0.55	&	7.17 $10^{10}$	\\
-1.25	&	5.73 $10^{11}$	\\
-2.00	&	1.11 $10^{12}$	\\

\hline
\end{tabular}

\label{tab:S2}
\end{table}

\newpage
\subsection{Measurements of brightening of dark complexes at varying gate voltages}
\begin{figure}[h]
		\subfloat{}%
		\centering
		\includegraphics[width=\linewidth]{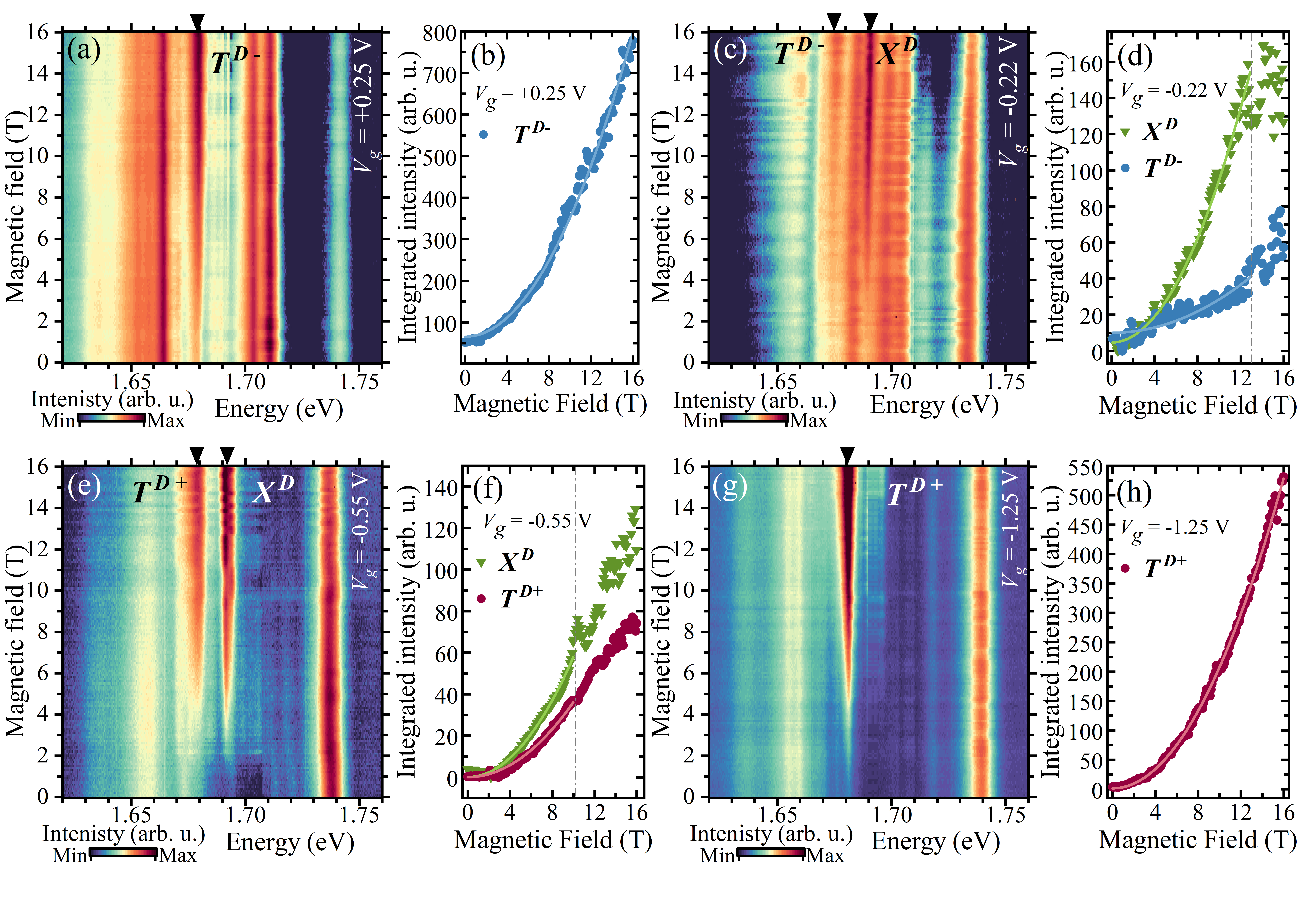}
        \caption{(a), (c), (e), (g) False-colour maps of low-temperature ($T = 10$~K) PL of a WSe$_2$ ML as a function of in-plane magnetic field for four gate voltages: (a) $V_g = +0.25$~V, (c) $V_g = -0.22$~V, (e) $V_g = -0.55$~V, and (g) $V_g = -1.25$~V. The colour-coded intensities in panels (a), (c), (e), and (g) are normalized to the maximum intensity in each panel for clarity. (b), (d), (f), (h) Magnetic-field dependence of the integrated intensities of dark excitonic complexes extracted for the four gate voltages: (b) $V_g = +0.25$~V, (d) $V_g = -0.22$~V, (f) $V_g = -0.55$~V, and (h) $V_g = -1.25$~V. For each dark excitonic complex, the corresponding quadratic fit is shown as a light-coloured line.}
		\label{fig:S2_maps}
\end{figure}

Figures~\ref{fig:S2_maps}(a), (c), (e), and (g) present false-colour maps of the low-temperature ($T = 10$~K) PL emission of the WSe$_2$ ML as a function of the in-plane magnetic field for four selected gate voltages: (a) $V_g = +0.25$~V, (c) $V_g = -0.22$~V, (e) $V_g = -0.55$~V, and (g) $V_g = -1.25$~V.
These gate voltages were chosen within the range explored in the main article.
In Figs.~\ref{fig:S2_maps}(a) and (g), the measurements are performed in the $n$-doped and $p$-doped regimes, respectively.
Accordingly, a pronounced increase in the intensity of the $T^{D-}$ and $T^{D+}$ lines is observed with increasing magnetic field.
In contrast, Figs.~\ref{fig:S2_maps}(c) and (e) correspond to conditions closer to the charge neutrality point.
In this case, an increase in the $X^D$ intensity with increasing magnetic field is clearly visible, accompanied by a weaker enhancement of the dark trion intensities.

To further investigate the brightening effect of these dark complexes, the PL spectra were deconvoluted using Lorentzian functions, following the procedure described in the main article.
The extracted evolution of the integrated intensities ($I_D$) of the $T^{D-}$, $X^D$, and $T^{D+}$ complexes as a function of the in-plane magnetic field is presented in Figs.~\ref{fig:S2_maps}(b), (d), (f), and (h) for the respective gate voltages.
These dependencies are expected to exhibit quadratic behaviour and can be described by $I_D = I_0 + \alpha B^2$, where $I_0$ is a fitting parameter corresponding to the zero-field intensity, and $\alpha$ denotes the brightening factor~\cite{Slobodeniuk2016, Molas_2017_DX_BDX, Zhang2017_DX, Molas2019_DX, Kipczak2024_DX_BDX_WSe2}.
As shown in Figs.~\ref{fig:S2_maps}(b), (d), (f), and (h), the experimental data are well reproduced by the fitted curves within the respective fitting ranges.

In Fig.~\ref{fig:S2_maps}(d), a spatial shift of the sample at 12~T results in an artificial change in the measured intensity.
Therefore, the fitting was restricted to magnetic fields up to 12~T.
A similar situation is observed in Fig.~\ref{fig:S2_maps}(e), where, above 9~T, anomalous spectral behaviour appears.
Specifically, the energy and intensity of the $X^D$ and $T^{D+}$ lines switch abruptly between two states, while the intensity continues to increase along two distinct trends.
This behaviour is likely caused by a slight spatial drift of the sample during the extended measurement time.
Consequently, the fitting in Fig.~\ref{fig:S2_maps}(f) was limited to magnetic fields below 9~T.
The extracted brightening factors are summarised in Table~\ref{tab:S1}.

\begin{table}[t]
\centering
\caption{The extracted brightening factors obtained from fitting the $I_D = I_0 + \alpha B^2$ dependence, corresponding to the data presented in Fig.~4 and Fig.~\ref{fig:S2_maps}(b), (d), (f), and (h).}
\begin{tabular}{|c|c|c|}
\hline
$V_g$ (V) & Excitonic complex & $\alpha$ (T$^{-2}$)\\
\hline
+1 & $T_{D-}$ & $3.507\pm0.035$ \\
+0.25 & $T_{D-}$ & $2.856\pm0.019$ \\
-0.22 & $T_{D-}$ & $0.1846\pm0.0072$ \\
-0.22 & $X_D$ & $0.9001\pm0.0010$ \\
-0.45 & $X_D$ & $1.236\pm0.016$ \\
-0.45 & $T_D$ & $347.8\pm8.3$ \\
-0.55 & $X_D$ & $0.5772\pm0.0053$ \\
-0.55 & $T_{D+}$ & $0.3605\pm0.0017$ \\
-1.25 & $T_{D+}$ & $2.0586\pm0.0065$ \\
-2 & $T_{D+}$ &$ 2.160\pm0.011$ \\
\hline
\end{tabular}

\label{tab:S1}
\end{table}

\end{document}